\documentclass[floatfix,showpacs,11pt,nofootinbib]{revtex4}
\usepackage{amssymb,amsmath}
\usepackage{graphicx,float}
\usepackage{indentfirst}

\newcommand{\beq}{\begin{equation}}
\newcommand{\eeq}{\end{equation}}
\newcommand{\bea}{\begin{eqnarray}}
\newcommand{\eea}{\end{eqnarray}}

\def\({\left(}
\def\){\right)}

\input{tcilatex}
\begin{document}

\title{ $f(\mathcal{R})$-Einstein-Palatini Formalism and smooth branes}
\author{P. Michel L. T. da Silva}
\email{pmichel@fc.unesp.br}
\affiliation{Departamento de F\'{\i}sica e Qu\'{\i}mica, Universidade Estadual Paulista,
Guaratinguet\'{a}, SP, Brazil}
\author{J. M. Hoff da Silva}
\email{hoff@feg.unesp.br}
\affiliation{Departamento de F\'{\i}sica e Qu\'{\i}mica, Universidade Estadual Paulista,
Guaratinguet\'{a}, SP, Brazil}

\begin{abstract}
In this work, we present the $f(\mathcal{R})$-Einstein-Palatini formalism in
arbitrary dimensions and the study of consistency applied to brane models,
the so-called braneworld sum rules. We show that it is possible a scenario
of thick branes in five dimensions with compact extra dimension in the
framework of the $f(R)$-Einstein-Palatini theory by the accomplishment of an
assertive criteria.
\end{abstract}

\maketitle

\section{Introduction}

Braneworld models have received a large amount of attention in the
high-energy community since the outstanding Randall-Sundrum model \cite{RS1}%
, providing a precise relation between a warped geometry and the mass scale
of an effective TeV universe. Soon after the establishment of warped models,
a plethora of models, generalizations, and applications where developed \cite%
{Maart}. Most importantly to our purposes was the smooth extension of warped
branes, first introduced by Gremm in \cite{Gremm}. From the perspective that
there must exist a typical length scale below what our understanding of the
physical laws should be, at lest, superseded by a full quantum gravity
theory, the idea of infinitely thin branes, as used in the Randall-Sundrum
model, is only an approximation, though highly nontrivial.

A crucial point concerning smooth extensions of branewords (see \cite{rev}
for a comprehensive review) in General Relativity theory is that it is
always necessary to preclude of the extra dimension orbifold topology used
in the original Randall-Sundrum model. After all, the $S^1/\mathcal{Z}_2$ is
also important to make contact to Ho\v{r}ava-Witten theory \cite{HW}. With
effect, there is an exhaustive theorem which forbids smooth generalizations
of the usual Randall-Sundrum model \cite{TEO} (see also \cite{LEB}). By
usual, we mean a five dimensional braneworld endowed with non-separable
geometry whose extra dimensions are compact, within the context of General
Relativity. In a gravitational theory different from General Relativity,
however, the situation may be different. In fact, by applying the so-called
braneworld sum rules (a set of consistency conditions obtained from the
gravitational equations of motion) it is possible to see that smooth
generalizations of the Randall-Sundrum framework are indeed possible.

Many extensions based upon the ideas delineated in the previous paragraph
was done. The investigation of braneworld sum rules applied to smooth branes
generalization in the context of Brans-Dicke and $f(\mathcal{R})$ gravity
has been studied in some detail in \cite{NS}. In these cases it is always
possible to show that the sum rules can be relaxed by the presence of
additional terms coming from the gravitational theory (other than the usual
case) in question. The $f(\mathcal{R})$ theory analyzed in \cite{NS} was
worked out in the light of the metric formalism. As it is well known,
however, that the metric and Palatini formalisms are not equivalent in the
approach to $f(\mathcal{R})$ gravity \cite{Tremblay,va}. One of the main
differences between the two approaches is given by the fact that in the
metric formalism the trace of fields equations gives rise to a dynamical
degree of freedom, whilst in the Palatini formalism this procedure lead to
an algebraic constraint. Concerning the problem we are interested here, we
shall see that the necessary condition leading to a smooth brane extension
is considerably modified, presenting a more clear criteria, namely, a $f(%
\mathcal{R})$ theory with negative first derivative with respect to $R$.

One of the difficulties that may arise when performing the sum rules regards
the Einstein tensor managing. Here, we use a simple conform transformation
in the metric to accomplish the sum rules program and extend the consistency
conditions in the scenario of theories $f(\mathcal{R})$ adopting the
Palatini formalism. This work is organized as follows. In Section II, we
briefly present the sum rules idea for braneworld scenarios. In the
following Section we construct the field equations in a $f(\mathcal{R})$%
-Einstein-Palatini formalism in arbitrary dimensions. In Section IV, we
apply the sum rules to the $f(\mathcal{R})$-Einstein-Palatini case,
investigating the relevant condition which leads to smooth branewords. In
the last Section we conclude.

\section{Sum rules for braneworld scenarios}

Much of the necessary formalism to the implementation of sum rules in the $f(%
\mathcal{R})$-Einstein-Palatini context was developed elsewhere \cite{LEB,NS}%
. Therefore, we shall pinpoint some important aspects in this Section. By
considering the spacetime as a $D$-dimensional manifold endowed with a
non-factorisable geometry, we write the line element as 
\begin{equation}
ds^{2}=g_{AB}(X)dX^{A}dX^{B}=W^{2}(r)g_{\mu \nu }(x)dx^{\mu }dx^{\nu
}+g_{ab}(r)dr^{a}dr^{b},
\end{equation}%
where $W^{2}(r)$ is the warp factor, $X^{A}$ denotes the coordinates of the
full $D$-dimensional spacetime, $x^{\mu }$ stands for the $(p+1)$
coordinates of the non-compact spacetime (brane), and $r^{a}$ labels the $%
(D-p-1)$ directions in the internal compact space. The classical action
takes into account the spacetime dynamics coupled to a scalar field, namely 
\begin{equation}
S=S_{gravity}+\int d^{D}X\,\sqrt{-g}\left( -\frac{1}{2}\partial _{A}\Phi
\partial ^{A}\Phi -V(\Phi )\right) ,
\end{equation}%
where we assume that the scalar field has only dependence on the internal
space coordinates $\Phi =\Phi (r^{m})$. The scalar field above shall be
understood as the responsible to generate the brane. We leave the potential
unspecified since it will not be relevant in our case. The energy-momentum
tensor gives 
\begin{equation}
T_{\mu \nu }=-W^{2}g_{\mu \nu }\left( \frac{1}{2}\nabla \Phi \cdot \nabla
\Phi +V(\Phi )\right) ,  \label{Ta}
\end{equation}%
and 
\begin{equation}
T_{ab}=\nabla _{a}\Phi \nabla _{b}\Phi -g_{ab}\left( \frac{1}{2}\nabla \Phi
\cdot \nabla \Phi +V(\Phi )\right) .  \label{Tb}
\end{equation}%
\indent 

It is possible to show \cite{TEO,LEB} that the following expression holds 
\begin{equation}
\nabla \cdot (W^{\alpha }\nabla W)=\frac{W^{\alpha +1}}{p(p+1)}\left[ \alpha
\,\left( W^{-2}\bar{R}-R_{\mu }^{\mu }\right) +(p-\alpha )\left( \tilde{R}%
-R_{a}^{a}\right) \right] ,  \label{consistency2}
\end{equation}%
where $R_{\mu }^{\mu }=W^{-2}g^{\mu \nu }R_{\mu \nu }$ and $%
R_{a}^{a}=g^{ab}R_{ab}$ are the partial traces such that $R=R_{\mu }^{\mu
}+R_{a}^{a}$ and $\alpha $ is an arbitrary parameter. Moreover, $\bar{R}$ is
the scalar of curvature derived from $g_{\mu \nu }$ and $\tilde{R}$ the
scalar of curvature associated to the internal space. The braneworld sum
rules can be obtained from two considerations, one physical and one
mathematical. From the physical point of view it is necessary to specify the
gravitational theory in question, i. e. write $S_{gravity}$. This being done
(notice that the dynamics is specified accordingly), one is able to use the
fact that, as far as the internal space is periodic without boundary, the
left hand side of (\ref{consistency2}) vanish under integration.

\section{The $f(R)$-Einstein-Palatini Formalism in Arbitrary Dimensions}

In the so-called Palatini formalism the metric and the connection are
assumed to be independent variables. The field equations are derived from
the variation of the Einstein-Hilbert action with respect to metric and
connection independently. Thus, the Ricci and Riemann tensors are objects
constructed from a general affine connection, but without the torsion terms.

It is well known that the definition $T_{AB}\equiv 2/\sqrt{-g}\delta
S_{M}/\delta g^{AB}$ when implemented along with the principle of least
action for $f(\mathcal{R})$-Einstein-Palatini gravity, leads to the
following field equations 
\begin{equation}
f^{\prime }(\mathcal{R})\mathcal{R}_{AB}-\dfrac{1}{2}f(\mathcal{R}%
)g_{AB}=8\pi G_{D}T_{AB},  \label{g1}
\end{equation}
and 
\begin{equation}
-\overset{\_}{\nabla }_{C}(\sqrt{-g}f^{\prime }(\mathcal{R})g^{AB})+\overset{%
\_}{\nabla }_{D}(\sqrt{-g}f^{\prime }(\mathcal{R})g^{D(A})\delta _{C}^{B)}=0,
\label{g2}
\end{equation}%
such that when $f(R)=$ $R$, the Palatini formalism restores general
relativity. Rewriting Eq. (\ref{g2}) we get 
\begin{equation}
-\overset{\_}{\nabla }_{C}(\sqrt{-g}f^{\prime }(\mathcal{R})g^{AB})+\text{ }%
\frac{1}{2}\text{ }\left[ \overset{\_}{\nabla }_{D}(\sqrt{-g}f^{\prime }(%
\mathcal{R})g^{DA})\delta _{C}^{B}+\overset{\_}{\nabla }_{D}(\sqrt{-g}%
f^{\prime }(\mathcal{R})g^{DB})\delta _{C}^{A}\right] =0,  \label{g3}
\end{equation}
and contracting the indices $C$ e $B$ we are left with 
\begin{equation}
\overset{\_}{\nabla }_{D}(\sqrt{-g}f^{\prime }(\mathcal{R})g^{DA})=0.
\label{g5}
\end{equation}
Therefore Eq. (\ref{g2}) reads simply 
\begin{equation}
\overset{\_}{\nabla }_{C}(\sqrt{-g}f^{\prime }(\mathcal{R})g^{AB})=0.
\label{g5b}
\end{equation}
In this vein, by defining a metric $h_{AB}$ as 
\begin{equation}
h_{AB}\equiv f^{\prime }(\mathcal{R})^{\frac{2}{D-2}}g_{AB},\text{ \ \ \ \ \
\ }h^{AB}\equiv f^{\prime }(\mathcal{R})^{\frac{2}{2-D}}g^{AB}\text{\ ,}
\label{g6}
\end{equation}
we formally have the connection equation 
\begin{equation}
\overset{\_}{\nabla }_{C}\left( \sqrt{-h}h^{AB}\right) =0.  \label{conex}
\end{equation}
Following this clue it is possible to write the connection as 
\begin{equation}
\overset{\_}{\Gamma }_{AB}^{\text{ \ \ \ }C}=\left\{ _{AB}^{\text{ \ \ \ }%
C}\right\} +\frac{1}{2f^{\prime }(\mathcal{R})^{\frac{2}{D-2}}}\Delta _{AB}^{%
\text{ \ \ \ }C},  \label{g8b}
\end{equation}
where 
\begin{equation}
\Delta _{AB}^{\text{ \ \ \ }C}=\left\{ \delta _{B}^{C}\partial _{A}f^{\prime
}(\mathcal{R})^{\frac{2}{D-2}}+\delta _{A}^{C}\partial _{B}f^{\prime }(%
\mathcal{R})^{\frac{2}{D-2}}-g_{AB}g^{CD}\partial _{D\text{ }}f^{\prime }(%
\mathcal{R})^{\frac{2}{D-2}}\right\} ,  \label{g8a}
\end{equation}
and $\left\{ _{AB}^{\text{ \ \ \ }C}\right\} $ are the usual Christoffel
symbols.

The Ricci tensor, generalized via the conformal (\ref{g6}) relation, is
given by $\mathcal{R}_{AB}=\partial _{C}\overset{\_}{\Gamma }_{AB}^{\text{ \
\ \ }C}-\partial _{B}\overset{\_}{\Gamma }_{AC}^{\text{ \ \ \ }C}+\overset{\_%
}{\Gamma }_{CE}^{\text{ \ \ \ }C}\overset{\_}{\Gamma }_{AB}^{\text{ \ \ \ }%
E}-\overset{\_}{\Gamma }_{BE}^{\text{ \ \ \ }C}\overset{\_}{\Gamma }_{AC}^{%
\text{ \ \ \ }E}$, and can be recast as 
\begin{equation}
\mathcal{R}_{AB}=R_{AB}+\dfrac{[D-1]}{2}\dfrac{\left( \nabla _{A}f^{\prime }(%
\mathcal{R})^{\frac{2}{D-2}}\right) \left( \nabla _{B}f^{\prime }(\mathcal{R}%
)^{\frac{2}{D-2}}\right) }{f^{\prime }(\mathcal{R})^{\frac{4}{D-2}}}-\dfrac{1%
}{f^{\prime }(\mathcal{R})^{\frac{2}{D-2}}}\left( \nabla _{A}\nabla _{B}+%
\frac{1}{2}g_{AB}\square \right) f^{\prime }(\mathcal{R})^{\frac{2}{D-2}},
\label{g9c}
\end{equation}%
and thus the generalized scalar of curvature reads 
\begin{equation}
\mathcal{R}\text{ }=R+\dfrac{[D-1]}{2}\dfrac{1}{f^{\prime }(\mathcal{R})^{%
\frac{4}{D-2}}}\left( \nabla _{A}f^{\prime }(\mathcal{R})^{\frac{2}{D-2}%
}\right) \left( \nabla ^{A}f^{\prime }(\mathcal{R})^{\frac{2}{D-2}}\right) -%
\dfrac{1}{f^{\prime }(\mathcal{R})^{\frac{2}{D-2}}}\left( \frac{D}{2}%
+1\right) \square f^{\prime }(\mathcal{R})^{\frac{2}{D-2}}.  \label{g10c}
\end{equation}%
In the Palatini formalism the field equations are given by%
\begin{equation}
\mathcal{R}_{AB}-\dfrac{f}{2f^{\prime }(\mathcal{R})}g_{AB}=\dfrac{8\pi
G_{D}T_{AB}}{f^{\prime }(\mathcal{R})}.  \label{P1}
\end{equation}%
Hence, inserting equation (\ref{g9c}) in (\ref{P1}) and adding on both sides
of the term $-g_{AB}R/2$ we obtain, after some manipulation, the
Einstein-Palatini field equations in arbitrary dimensions 
\begin{gather}
R_{AB}-\dfrac{1}{2}Rg_{AB}=\dfrac{8\pi G_{D}T_{AB}}{F(f^{\prime }(\mathcal{R}%
))}-\dfrac{g_{AB}}{2}\left( \mathcal{R}-\dfrac{f(\mathcal{R})}{F(\mathcal{R})%
}\right) +\dfrac{1}{F(\mathcal{R})^{\frac{2}{D-2}}}\left( \nabla _{A}\nabla
_{B}-g_{AB}\square \right) F(\mathcal{R})^{\frac{2}{D-2}}  \notag \\
-\dfrac{[D-1]}{2F(\mathcal{R})^{\frac{4}{D-2}}}\left[ \left( \nabla _{A}F(%
\mathcal{R})^{\frac{2}{D-2}}\right) \left( \nabla _{B}F(\mathcal{R})^{\frac{2%
}{D-2}}\right) -\dfrac{g_{AB}}{2}\nabla _{C}F(\mathcal{R})^{\frac{2}{D-2}%
}\nabla ^{C}F(\mathcal{R})^{\frac{2}{D-2}}\right] .  \label{P3}
\end{gather}%
where $F(\mathcal{R})=df(\mathcal{R})/d\mathcal{R}$ and $\mathcal{R}$ is
Ricci scalar constructed out from $\mathcal{R}_{AB}$. Now we are able to
implement the relevant partial traces, derived from (\ref{P3}), into Eq. (%
\ref{consistency2}).

\section{Braneword sum rules in $f(R)$-Einstein-Palatini}

Taking advantage of Eq. (\ref{P3}) we see that the scalar of curvature reads 
\begin{eqnarray}
R &=&\dfrac{2}{(2-D)}\left\{ \dfrac{8\pi G_{D}}{F(\mathcal{R})}T-\dfrac{D}{2}%
\left( \mathcal{R}-\dfrac{f(\mathcal{R})}{F(\mathcal{R})}\right) -(D-1)%
\dfrac{1}{F(\mathcal{R})^{\frac{2}{D-2}}}\square F(\mathcal{R})^{\frac{2}{D-2%
}}\right.  \notag \\
&&\left. -\dfrac{[D-1]}{2F(\mathcal{R})^{\frac{4}{D-2}}}\left[ \left( \nabla
_{A}F(\mathcal{R})^{\frac{2}{D-2}}\right) \left( \nabla ^{A}F(\mathcal{R})^{%
\frac{2}{D-2}}\right) -\dfrac{D}{2}\nabla _{C}F(\mathcal{R})^{\frac{2}{D-2}%
}\nabla ^{C}F(\mathcal{R})^{\frac{2}{D-2}}\right] \right\},  \label{g12c}
\end{eqnarray}
from which, reinserting it back in (\ref{P3}), we have 
\begin{eqnarray}
R_{AB} &=&\frac{1}{F(\mathcal{R})}\left[ 8\pi G_{D}\left( T_{AB}-\dfrac{%
g_{AB}}{(D-2)}T\right) \right] +\frac{\nabla _{A}\nabla _{B}F(\mathcal{R})^{%
\frac{2}{D-2}}}{F(\mathcal{R})^{\frac{2}{D-2}}}+\dfrac{g_{AB}}{(D-2)}\left\{
\left( \mathcal{R}-\dfrac{f(\mathcal{R})}{F(\mathcal{R})}\right) +\dfrac{%
\square F(\mathcal{R})^{\frac{2}{D-2}}}{F(\mathcal{R})^{\frac{2}{D-2}}}%
\right.  \notag \\
&&\left. -\frac{\nabla _{C}F(\mathcal{R})^{\frac{2}{D-2}}\nabla ^{C}F(%
\mathcal{R})}{2F(\mathcal{R})^{\frac{4}{D-2}}}\right\} -\frac{(D-1)(D-3)}{%
2(D-2)F(\mathcal{R})^{\frac{4}{D-2}}}\left( \nabla _{A}F(\mathcal{R})^{\frac{%
2}{D-2}}\right) \left( \nabla _{B}F(\mathcal{R})^{\frac{2}{D-2}}\right).
\label{13}
\end{eqnarray}
The partial trace of the above equation with respect to the brane,
non-compact, dimensions is given by 
\begin{eqnarray}
R_{\mu }^{\mu } &=&\left. \frac{1}{(D-2)F(\mathcal{R})}\left[ 8\pi G_{D}%
\bigg((D-p-3)T_{\mu }^{\mu }-(p+1)T_{a}^{a}\bigg)\right] \right. +\dfrac{%
(D+p-1)}{(D-2)F(\mathcal{R})^{\frac{2}{D-2}}}{\LARGE (}W^{-2}\nabla _{\mu
}\nabla ^{\mu }F(\mathcal{R})^{\frac{2}{D-2}}{\LARGE )}  \notag \\
&&+\dfrac{(p+1)}{(D-2)}\left[ \left( \mathcal{R}-\dfrac{f(\mathcal{R})}{F(%
\mathcal{R})}\right) +\dfrac{\nabla _{a}\nabla ^{a}F(R)^{\frac{2}{D-2}}}{F(%
\mathcal{R})^{\frac{2}{D-2}}}-\frac{1}{2F(R)^{\frac{4}{D-2}}}\bigg({\LARGE (}%
W^{-2}\nabla _{\lambda }F(R)^{\frac{2}{D-2}}\nabla ^{\lambda }F(R)^{\frac{2}{%
D-2}}\right.  \notag \\
&&\left. +\nabla _{c}F(\mathcal{R})^{\frac{2}{D-2}}\nabla ^{c}F(\mathcal{R}%
)^{\frac{2}{D-2}}{\LARGE )}\bigg)\right] -\frac{(D-1)(D-3)}{2(D-2)F(\mathcal{%
R})^{\frac{4}{D-2}}}W^{-2}\left( \nabla _{\mu }F(\mathcal{R})^{\frac{2}{D-2}%
}\right) \left( \nabla ^{\mu }F(\mathcal{R})^{\frac{2}{D-2}}\right) ,
\label{14}
\end{eqnarray}
while its internal space counterpart reads 
\begin{eqnarray}
R_{a}^{a} &=&\frac{1}{(D-2)F(\mathcal{R})}\left[ 8\pi G_{D}\bigg(%
(p-1)T_{a}^{a}-(D-p-1)T_{\mu }^{\mu }\bigg)\right] +\dfrac{(2D-p-3)}{F(%
\mathcal{R})^{\frac{2}{D-2}}(D-2)}{\LARGE (}W^{-2}\nabla _{\mu }\nabla ^{\mu
}F(\mathcal{R})^{\frac{2}{D-2}}{\LARGE )}  \notag \\
&&+\dfrac{(D-p-1)}{(D-2)}\Bigg[\left( \mathcal{R}-\dfrac{f(\mathcal{R})}{F(%
\mathcal{R})}\right) +\dfrac{{\LARGE (}\nabla _{a}\nabla ^{a}F(\mathcal{R})^{%
\frac{2}{D-2}}{\LARGE )}}{F(\mathcal{R})^{\frac{2}{D-2}}}-\dfrac{1}{2F(%
\mathcal{R})^{\frac{4}{D-2}}}{\LARGE (}W^{-2}\nabla _{\lambda }F(\mathcal{R}%
)^{\frac{2}{D-2}}\nabla ^{\lambda }F(\mathcal{R})^{\frac{2}{D-2}}  \notag \\
&&+\nabla _{c}F(\mathcal{R})^{\frac{2}{D-2}}\nabla ^{c}F(\mathcal{R})^{\frac{%
2}{D-2}}{\LARGE )}\Bigg]-\frac{(D-1)(D-3)}{2(D-2)F(\mathcal{R})^{\frac{4}{D-2%
}}}\left( \nabla _{a}F(\mathcal{R})^{\frac{2}{D-2}}\right) \left( \nabla
^{a}F(\mathcal{R})^{\frac{2}{D-2}}\right) .  \label{15}
\end{eqnarray}
Now, by inserting (\ref{14}) and (\ref{15}) into equation (\ref{consistency2}%
), one arrives at 
\begin{eqnarray}
\nabla \cdot (W^{\alpha }\nabla W) &=&\frac{W^{\alpha +1}}{p(p+1)(D-2)F(%
\mathcal{R})}\bigg\{8\pi G_{D}\bigg((p-\alpha )(D-p-1)-\alpha (D-p-3)\bigg)%
T_{\mu }^{\mu }+  \notag \\
&&+\,8\pi G_{D}\bigg(\alpha (p+1)-(p-\alpha )(p-1)\bigg)T_{a}^{a}+(D-2)\bigg(%
\alpha W^{-2}\bar{R}+(p-\alpha )\tilde{R}\bigg)F(\mathcal{R})-  \notag \\
&&-\frac{W^{\alpha +1}}{p(p+1)(D-2)}\left\{ \left[ \dfrac{W^{-2}\nabla _{\mu
}\nabla ^{\mu }F(\mathcal{R})^{\frac{2}{D-2}}}{F(\mathcal{R})^{\frac{2}{D-2}}%
}\right] \left[ \alpha (D+p-1)+(p-\alpha )(2D-p-3)\right] \right.  \notag \\
&&-\left[ \alpha (p+1)+(p-\alpha )(D-p-1)\right] \left[ \left( \mathcal{R}-%
\dfrac{f(\mathcal{R})}{F(\mathcal{R})}\right) +\dfrac{{\LARGE (}\nabla
_{a}\nabla ^{a}F(\mathcal{R})^{\frac{2}{D-2}}{\LARGE )}}{F(\mathcal{R})^{%
\frac{2}{D-2}}}\right.  \notag \\
&&\left. -\dfrac{1}{F(\mathcal{R})^{\frac{4}{D-2}}}{\LARGE (}W^{-2}\nabla
_{\lambda }F(\mathcal{R})^{\frac{2}{D-2}}\nabla ^{\lambda }F(\mathcal{R})^{%
\frac{2}{D-2}}+\nabla _{c}F(\mathcal{R})^{\frac{2}{D-2}}\nabla ^{c}F(%
\mathcal{R})^{\frac{2}{D-2}}{\LARGE )}\right]  \notag \\
&&+\left[ \frac{(D-1)(D-3)}{2F(\mathcal{R})^{\frac{4}{D-2}}}\right] \left[
\alpha (W^{-2}\nabla _{\lambda }F(\mathcal{R})^{\frac{2}{D-2}}\nabla
^{\lambda }F(\mathcal{R})^{\frac{2}{D-2}})+(p-\alpha )\right. ,  \notag \\
&&\left. \left. \times \left( \nabla _{a}F(\mathcal{R})^{\frac{2}{D-2}%
}\right) \left( \nabla ^{a}F(\mathcal{R})^{\frac{2}{D-2}}\right) \right]
\right\} .  \label{eit}
\end{eqnarray}

As a last step, by assuming the internal space compact (as in the standard
cases) the left hand side of Eq. (\ref{eit}) vanishes upon integration.
Following the standard presentation we denote these integrations by $\oint
\nabla \cdot (W^{\alpha}\nabla W)=0$. Hence, by inserting the
energy-momentum partial traces and integrating over the internal space, it
is possible to obtain the sum rules to the very general case in the scope of 
$f(\mathcal{R})$-Eintein-Palatini theory. The result is quite large, and its
generality contributes to overshadow its physical content.

In order to extract physical information it is convenient to particularize
the analysis to the $D=5$ and $p=3$ case. Thus we shall investigate a
five-dimensional bulk with an unique extra dimension ($\tilde{R}=0$) endowed
to a orbifold topology, for instance. Besides, in this four-dimensional
brane context we can implement the physical constraint $(\bar{R}=0)$ in
trying to describe our universe in large scales. Therefore, after these
particularizations, and using equations (\ref{Ta}) and (\ref{Tb}) for the
sources, we have the following set of conditions 
\begin{eqnarray}
&0 &\,=\,8\pi G_{5}\oint \frac{W^{\alpha +1}}{F(\mathcal{R})}\bigg\{%
(3-\alpha )\Phi ^{\prime }\cdot \Phi ^{\prime }+2(\alpha +1)V(\Phi )\bigg\} 
\notag \\
&&+(\alpha +1)\bigg\{4\oint \dfrac{W^{^{\alpha -1}}\nabla _{\mu }\nabla
^{\mu }F(\mathcal{R})^{2/3}}{F(\mathcal{R})^{2/3}}+\oint W^{^{\alpha +1}}%
\left[ \left( \mathcal{R}-\dfrac{f(\mathcal{R})}{F(\mathcal{R})}\right) +%
\dfrac{{\LARGE (}\nabla _{a}\nabla ^{a}F(\mathcal{R})^{2/3}{\LARGE )}}{F(%
\mathcal{R})^{2/3}}\right]  \notag \\
&&-\dfrac{1}{6}\oint \dfrac{W^{\alpha -1}}{F(\mathcal{R})^{4/3}}\nabla _{\mu
}F(\mathcal{R})^{2/3}\nabla ^{\mu }F(\mathcal{R})^{2/3}+\dfrac{1}{6}\oint 
\dfrac{W^{\alpha -1}}{F(\mathcal{R})^{4/3}}\nabla _{a}F(\mathcal{R}%
)^{2/3}\nabla ^{a}F(\mathcal{R})^{2/3}\bigg\}  \notag \\
&&+\dfrac{4}{3}\alpha\!\oint\! \dfrac{W^{\alpha -1}}{F(\mathcal{R})^{4/3}}%
\nabla _{\mu }F(\mathcal{R})^{2/3}\nabla ^{\mu }F(\mathcal{R})^{2/3}+\frac{4%
}{3}(3-\alpha )\!\oint\! \dfrac{W^{\alpha +1}}{F(\mathcal{R})^{4/3}}\nabla
_{a}F(\mathcal{R})^{2/3}\nabla ^{a}F(\mathcal{R})^{2/3}.  \label{ke}
\end{eqnarray}

Among all consistency conditions encoded in (\ref{ke}), each related to a
given $\alpha$, there are many irrelevant. As a matter of fact, in order to
explore the smooth branes possibility the choice $\alpha=-1$ is particularly
elucidative, since it eliminates the overall warp factor. In fact, this
choice provides simply 
\begin{equation}
\oint \frac{\Phi ^{\prime }\cdot \Phi ^{\prime }}{F(\mathcal{R})}+\dfrac{1}{%
6\pi G_{5}}\oint \dfrac{\nabla _{a}F(\mathcal{R})^{2/3}\nabla ^{a}F(\mathcal{%
R})^{2/3}}{F(\mathcal{R})^{4/3}}=0,  \label{ul}
\end{equation}
in which $\nabla _{\mu }F(R)^{2/3}=0$ was already taken into account.
Interestingly enough, Eq. (\ref{ul}) may be rewritten as 
\begin{equation}
\oint \frac{\Phi ^{\prime }\cdot \Phi ^{\prime }}{F(\mathcal{R})}+\dfrac{1}{%
27\pi G_{5}}\oint \left( \ln \left\vert F(\mathcal{R})\right\vert \right)
^{\prime }\cdot\left( \ln \left\vert F(\mathcal{R})\right\vert \right)
^{\prime }=0.
\end{equation}
Now it turns out that whether $F(\mathcal{R})$ is positive, then it is
impossible to achieve a smooth generalization of usual braneworld models,
since the resulting constraint 
\begin{equation}
\oint \Bigg(\frac{1}{F(\mathcal{R})^{1/2}}\frac{d\Phi}{dr}\Bigg)\cdot\Bigg(%
\frac{1}{F(\mathcal{R})^{1/2}}\frac{d\Phi}{dr}\Bigg) +\dfrac{1}{27\pi G_{5}}%
\oint \left( \ln \left\vert F(\mathcal{R})\right\vert \right) ^{\prime
}\cdot\left( \ln \left\vert F(\mathcal{R})\right\vert \right) ^{\prime }=0,
\end{equation}
can never be satisfied. The situation is utterly different in the case of a
negative $F(\mathcal{R})$. Obviously, in this last case the balance relation 
\begin{equation}
\oint \Bigg(\frac{1}{\vert F(\mathcal{R})\vert ^{1/2}}\frac{d\Phi}{dr}\Bigg)%
\cdot\Bigg(\frac{1}{\vert F(\mathcal{R})\vert ^{1/2}}\frac{d\Phi}{dr}\Bigg) =%
\dfrac{1}{27\pi G_{5}}\oint \left( \ln \left\vert F(\mathcal{R})\right\vert
\right) ^{\prime }\cdot\left( \ln \left\vert F(\mathcal{R})\right\vert
\right) ^{\prime},  \label{28}
\end{equation}
may be satisfied. Equation (\ref{28}) perform, then, a clear criteria -- $F(%
\mathcal{R}<0)$ -- for the possibility of smooth 3-branes in a five
dimensional bulk. In comparing with the metric approach, where the negative
quantity were proportional to $\oint F(\mathcal{R})^{-1}\nabla^2F(\mathcal{R}%
)$, the result obtained in the Palatini context is indeed exhaustive. As a
final remark, notice that in the limit $F(\mathcal{R})\rightarrow 1$, i. e. $%
f(\mathcal{R})\rightarrow \mathcal{R}$, Eq. (\ref{28}) reduces to $\oint
\Phi^{\prime}\cdot\Phi^{\prime}=0$, just as in usual General Relativity (as
expected), a constraint which can never be reached.

\section{Concluding Remarks}

The modeling of warped smooth branes has given rise to a somewhat more
formal branch of research in the context of braneworld gravity. In turn,
this line of investigation has lead to the solidification of braneworld
models in several different perspectives, since non-compact extra dimension 
\cite{ahm}, different bulk cosmological constants \cite{ahm2}, and ingenious
single thick branes approach \cite{ahm3}, just to enumerate some. In which
concern this paper, the general idea is not to set a specific model, but
instead to provide a comprehensive scope from which consistent models can be
constructed up.

It is shown that smooth generalizations of the usual Randall-Sundrum
braneworld model can be achieved in $f(\mathcal{R})$ gravity. This is
already know from previous work \cite{NS}, but here we have worked in the
Palatini formalism to $f(\mathcal{R})$. Apart from the fact that the metric
and the Palatini formalisms to $f(\mathcal{R})$ are inequivalent, the
analysis performed here has culminating into a more clear and assertive
constraint to be fulfilled.

Even though it is not our purpose here the proposition of models, we shall
emphasize that among all the possible generalizations leading to smooth
3-branes in a five dimensional within non-separable geometry (and a compact
extra dimension), the use of $f(\mathcal{R})$-Eintein-Palatini formalism
seems to be a quite promising approach. This is because, once again, in this
context it is possible to extract a simple and sharp necessary criteria.

\section*{Acknowledgments}

It is a pleasure to thank Prof. A. de Souza Dutra and T. R. P. Carames for
useful conversation. PMLTS acknowledge CAPES for financial support and JMHS
thanks to CNPq (308623/2012-6; 445385/2014-6) for financial support.

\end{document}